\documentclass[fleqn,10pt]{article}
\usepackage{cite}
\usepackage[pdftex]{hyperref}
\usepackage{amsmath, amsthm, amssymb, graphicx,slashed,mathrsfs}
\parskip=\baselineskip

\font\teneurm=eurm10 \font\seveneurm=eurm7 \font\fiveeurm=eurm5
\newfam\eurmfam
\textfont\eurmfam=\teneurm \scriptfont\eurmfam=\seveneurm
\scriptscriptfont\eurmfam=\fiveeurm

 \font\teneusm=eusm10 \font\seveneusm=eusm7 \font\fiveeusm=eusm5
\newfam\eusmfam
\textfont\eusmfam=\teneusm \scriptfont\eusmfam=\seveneusm
\scriptscriptfont\eusmfam=\fiveeusm

\font\tencmmib=cmmib10 \skewchar\tencmmib='177
\font\sevencmmib=cmmib7 \skewchar\sevencmmib='177
\font\fivecmmib=cmmib5 \skewchar\fivecmmib='177
\newfam\cmmibfam
\textfont\cmmibfam=\tencmmib \scriptfont\cmmibfam=\sevencmmib
\scriptscriptfont\cmmibfam=\fivecmmib

\parskip=\baselineskip

\def\p{\mbox{\boldmath$\displaystyle\mathbf{p}$}}

\def\bv{\mbox{\boldmath$\displaystyle\mathbf{\varphi}$}}

\def\0{\mbox{\boldmath$\displaystyle\mathbf{0}$}}

\def\K{\mbox{\boldmath$\displaystyle\mathbf{K}$}}
\def\x{\mbox{\boldmath$\displaystyle\mathbf{x}$}}

\begin{document}

\vskip 1.5in
\begin{center}
{\bf\Large{Massive fermions in 2+1 dimensions}}\vskip 0.5cm
{Cheng-Yang Lee} \vskip 0.1in
{\small{
{\textit{Institute of Mathematics, Statistics and Scientific Computation,\\
Unicamp, 13083-859 Campinas, S\~{a}o Paulo, Brazil.\\
Email: cylee@ime.unicamp.br}}}}
\end{center}

\begin{abstract}
We construct the higher-spin massive fermionic fields in 2+1 dimensions. Their field equations and propagators are derived from first principle. For fields with $j>\frac{1}{2}$, complications arise from the non-linear behaviour of the boost operators. We find that for a spin-three-half field, the non-linearity have an underlying structure that guarantees the locality of the fields and the existence of covariant propagators. We conjecture that this structure exists for fields of all spin.
\end{abstract}

\section{Introduction}


In this note, we construct higher-spin massive fermionic fields in 2+1 dimensions from first-principle. While the spin-half Dirac field is well-known and contains many fascinating properties, to the best of our knowledge, its higher-spin generalisation have not received much attention. The construction is based on the formalism developed by Wigner and Weinberg~\cite{Wigner:1939cj,Weinberg:1964cn,Weinberg:1964ev,Weinberg:1995mt} where the fermionic states and fields furnish irreducible representations of the Poincar\'{e} group. 

An important property for the massive particle states is that their little group is $SO(2)$ where all the irreducible unitary representations are one-dimensional. This means that for a spin-$j$ representation where $j$ is a half-integer, there exists $2j+1$ unique fermionic fields $\psi^{(\sigma)}(x)$ where $\sigma=-j,\cdots j$. Here their field equations and propagators are derived from first-principle. We show that because the representations are one-dimensional, the mass terms for the field equations and propagators are $\sigma$-dependent. As we shall see, for fermionic fields with $j>\frac{1}{2}$, complications arise due to the non-linear behaviour of the boost operators so that additional analysis and demands are required to determine the locality of the fields and their propagators.



\section{Massive particle states and fields}
In this section, we start with a review of the Poincar\'{e} algebra in 2+1 dimensions and its representations. Subsequently, the formalism for constructing massive fermionic fields is presented followed by the derivation of the Dirac equation. In 2+1 dimensions, the algebra can be compactly written as
\begin{subequations}
\begin{equation}
i[J^{\mu},J^{\nu}]=\epsilon^{\mu\nu\rho}J_{\rho},
\end{equation}
\begin{equation}
i[J^{\mu},P^{\nu}]=\epsilon^{\mu\nu\rho}P_{\rho},
\end{equation}
\begin{equation}
[P^{\mu},P^{\nu}]=0
\end{equation}
\end{subequations}
with $\mu=0,1,2$. The operator $J^{\mu}$ is defined as
\begin{equation}
J^{\mu}=\frac{1}{2}\epsilon^{\mu\nu\rho}J_{\nu\rho}
\end{equation} 
where $J^{\nu\rho}$ and $P^{\mu}$ are the generators of the Lorentz algebra and space-time translations respectively. The boosts and rotation generators are identified as
\begin{equation}
J^{01}=K_{x},\hspace{0.5cm}J^{02}=K_{y},\hspace{0.5cm} J^{12}=J_{z}.
\end{equation}
The space-time metric and the Levi-Civita tensor are chosen to be $\eta^{\mu\nu}=\mbox{diag}(1,-1,-1)$ and $\epsilon^{012}=1$ respectively. In terms of $J^{\mu}$ and $P^{\mu}$, the Casimir invariants are
\begin{equation}
C_{1}=P^{\mu}P_{\mu},\hspace{0.5cm}
C_{2}=-P_{\mu}J^{\mu}.
\end{equation}

As usual, the one-particle state $|p,\sigma\rangle$ is defined as an eigenstate of $P^{\mu}$
\begin{equation}
P^{\mu}|p,\sigma\rangle=p^{\mu}|p,\sigma\rangle
\end{equation}
where $\sigma$ is an arbitrary index that can be either continuous or discrete. The states must also be simultaneous eigenstates of the Casimir invariants.
To determine their spectrum for massive particle states and the index $\sigma$, it is convenient to start with the state at rest $|k,\sigma\rangle$ where $k^{\mu}=(m,\0)$. It satisfies
\begin{equation}
C_{1}|k,\sigma\rangle=m^{2}|k,\sigma\rangle.
\end{equation}
Under the action of $C_{2}$, we have
\begin{equation}
C_{2}|k,\sigma\rangle=-mJ_{z}|k,\sigma\rangle.
\end{equation}
Since the state are defined to be simultaneous eigenstates of $C_{1}$ and $C_{2}$, in the rest frame, $|k,\sigma\rangle$ must be an eigenstate of $J_{z}$
\begin{equation}
J_{z}|k,\sigma\rangle=s|k,\sigma\rangle \label{eq:jz}
\end{equation} 
thus giving us
\begin{equation}
C_{2}|k,\sigma\rangle=-ms|k,\sigma\rangle.
\end{equation}
Therefore, the spectrum of $C_{1}$ and $C_{2}$ are $m^{2}$ and $-ms$ respectively. It is important to note that the notion of spin defined by eq.~(\ref{eq:jz}) only holds in the rest frame since the boost generators do not commute with $J_{z}$. The index $\sigma$ and $s$ can now be determined by the irreducible unitary representations of the Lorentz group. Towards this end, we use the fact that the Lorentz transformation on a one-particle state $|p,\sigma\rangle$ is given by~\cite{Weinberg:1995mt}
\begin{equation}
U(\Lambda)|p,\sigma\rangle=\sqrt{\frac{E_{\Lambda p}}{E_{p}}}\sum_{\sigma'}D_{\sigma'\sigma}(W(\Lambda,p))|\Lambda p,\sigma'\rangle\label{eq:particle_trans}
\end{equation}
where the matrix $D(W(\Lambda,p))$ is the irreducible unitary representation of the little group. The elements of the little group are defined as
\begin{equation}
W(\Lambda,p)=L^{-1}(\Lambda p)\Lambda L(p) \label{eq:little}
\end{equation}
with $L(p)$ being the standard Lorentz boost taking particles at rest to arbitrary momentum. The normalisation factor in eq.~(\ref{eq:particle_trans}) is chosen so that the inner-product between particle states are
\begin{equation}
\langle p',\sigma'|p,\sigma\rangle=\delta_{\sigma'\sigma}\delta^{3}(\p'-\p).
\end{equation}

In 2+1 dimensions, the little group for the massive particle is $SO(2)$. Its multi-valued irreducible unitary representation is given by~\cite{Binegar:1981gv}
\begin{equation}
D_{\sigma'\sigma}(W(\Lambda,p))=e^{i\sigma\phi(\Lambda,p)}\delta_{\sigma'\sigma}\label{eq:massive_little_group}
\end{equation}
where $\sigma\in R$ is a continuous real number. We shall see shortly that for the massive fermions under consideration, $\sigma$ is restricted to half-integers.
Substituting eq. (\ref{eq:massive_little_group}) into eq. (\ref{eq:particle_trans}), we obtain
\begin{equation}
U(\Lambda)|p,\sigma\rangle=\sqrt{\frac{E_{\Lambda p}}{E_{p}}}e^{i\sigma\phi(\Lambda,p)}|\Lambda p,\sigma\rangle.\label{eq:state_transform}
\end{equation}
The angle $\phi(\Lambda,p)$ is the angle of rotation determined by eq.~(\ref{eq:little}). The computation of $W(\Lambda,p)$ in 3+1 dimensions can be found in~\cite{Ferraro:1999eu}. This can be adapted to the 2+1 dimensions case thus allowing us to determine $\phi(\Lambda,p)$. To determine $s$, we apply rotation about the 3-axis on the state $|k,\sigma\rangle$
\begin{equation}
U(R(\phi))|k,\sigma\rangle=e^{i\sigma\phi}|k,\sigma\rangle.
\end{equation}
Expand both sides about the identity we get
\begin{equation}
J_{z}|k,\sigma\rangle=\sigma|k,\sigma\rangle.
\end{equation}
Comparing this with eq.~(\ref{eq:jz}), we obtain $\sigma=s\in\mathbb{R}$.

Since $SO(2)$ is Abelian, its irreducible unitary representations are all one-dimensional. Therefore, the resulting quantum field operator associated with the state $|p,\sigma\rangle$ takes the form
\begin{equation}
\psi^{(\sigma)}(x)=(2\pi)^{-1}\int \frac{d^{2}p}{\sqrt{2E_{p}}}\Big[e^{-ip\cdot x}u(\p,\sigma)a(\p,\sigma)+e^{ip\cdot x}v(\p,\sigma)b^{\dag}(\p,\sigma)\Big]\label{eq:qf}
\end{equation}
where $u(\p,\sigma)$ and $v(\p,\sigma)$ are the expansion coefficients to be determined. The field $\psi^{(\sigma)}(x)$ is covariant under space-time translations. The demand of Lorentz-covariance means that it must transform as
\begin{equation}
U(\Lambda)\psi^{(\sigma)}(x)U^{-1}(\Lambda)= \mathcal{D}
(\Lambda^{-1})\psi^{(\sigma)}(\Lambda x)\label{eq:vsr_qf_transform}
\end{equation}
where $\mathcal{D}(\Lambda)$ is the finite-dimensional 
representation of the Lorentz group. The constraints on the expansion coefficients are obtained by substituting eq.~(\ref{eq:qf}) into (\ref{eq:vsr_qf_transform}) and using the Lorentz transformations of the annihilation and creation operator which can be obtained from eq.~(\ref{eq:state_transform}). After some algebraic manipulations, we obtain the following constraints
\begin{subequations}
\begin{equation}
u_{\bar{\ell}}(\mathbf{\Lambda}\p,\sigma)D(W(\Lambda,p))=\sum_{\ell}
\mathcal{D}_{\bar{\ell}\ell}(\Lambda)u_{\ell}(\p,\sigma),\label{eq:u_constraint}
\end{equation}
\begin{equation}
v_{\bar{\ell}}(\mathbf{\Lambda}\p,\sigma)D^{*}(W(\Lambda,p))=\sum_{\ell}
\mathcal{D}(\Lambda)v_{\ell}(\p,\sigma).\label{eq:v_constraint}
\end{equation}
\end{subequations}
We now solve eqs.~(\ref{eq:u_constraint}) and (\ref{eq:v_constraint}) to determine the expansion coefficients. First we take $p^{\mu}=k^{\mu}$ and $\Lambda=L(p)$. Using the identity $W(L(p),p)=I$, the expansion coefficients at rest and at arbitrary momentum are related by
\begin{subequations}
\begin{equation}
u(\p,\sigma)=\mathcal{D}(L(p))u(\0,\sigma),
\end{equation}
\begin{equation}
v(\p,\sigma)=\mathcal{D}(L(p))v(\0,\sigma).
\end{equation}
\end{subequations}
Here the boost operator is given by
\begin{equation}
\mathcal{D}(L(p))=\exp(-i\K\cdot\bv)
\end{equation}
where $\bv=\varphi\hat{\p}$ is the rapidity parameter defined as
\begin{equation}
\cosh\varphi=\frac{E_{p}}{m},\hspace{0.5cm}
\sinh\varphi=\frac{|\p|}{m}.
\end{equation}
The expansion coefficients at rest are determined by taking $p^{\mu}=k^{\mu}$ and $\Lambda=R(\phi)$, we get
\begin{subequations}
\begin{equation}
u_{\bar{\ell}}(\mathbf{k},\sigma)D(R(\phi))=\sum_{\ell}
\mathcal{D}_{\bar{\ell}\ell}(R(\phi))u_{\ell}(\mathbf{k},\sigma),
\end{equation}
\begin{equation}
v_{\bar{\ell}}(\mathbf{k},\sigma)D^{*}(R(\phi))=
\sum_{\ell}\mathcal{D}(R(\phi))_{\bar{\ell}\ell}v_{\ell}(\mathbf{k},\sigma)
\end{equation}
\end{subequations}
where we have used the identity $W(R(\phi),p)=R(\phi)$.
Expand both sides of the equations about the identity, we find that the rest coefficients are eigenvectors of the rotation generator
\begin{subequations}
\begin{equation}
\mathcal{J}_{z}u(\0,\sigma)=\sigma u(\0,\sigma),
\end{equation}
\begin{equation}
\mathcal{J}_{z}v(\0,\sigma)=-\sigma v(\0,\sigma).
\end{equation}
\end{subequations}
We take $\mathcal{J}_{z}$ to be a finite-dimensional matrix so the label $\sigma$ instead of being continuous, is now restricted to the eigenvalues of $\mathcal{J}_{z}$ 
and is therefore discrete. 

\subsection{The Dirac field}
We now have the necessary ingredients to construct massive fermionic fields in 2+1 dimensions. As a warm-up exercise, we construct the spin-half fields and then proceed to its higher-spin generalisation in the next section. To apply the above formalism, we need to construct finite-dimensional generators of the Lorentz group. This is achieved by introducing the following $\alpha$-matrices~\cite{Binegar:1981gv}
\begin{equation}
\alpha^{0}=\left(\begin{matrix}
1 & 0 \\
0 &-1 \end{matrix}\right),\hspace*{0.5cm}
\alpha^{1}=\left(\begin{matrix}
0 & -i\\
-i & 0
\end{matrix}\right),\hspace*{0.5cm}
\alpha^{2}=\left(\begin{matrix}
0 & -1\\
1& 0
\end{matrix}\right)\label{eq:two_component_gamma}
\end{equation}
where they satisfy  the Clifford algebra $\{\alpha^{\mu},\alpha^{\nu}\}=2\eta^{\mu\nu}I$. The generators of the Lorentz group are given by 
\begin{equation}
\mathcal{J}^{\mu\nu}=\frac{i}{4}[\alpha^{\mu},\alpha^{\nu}].\label{eq:two_component_J}
\end{equation}
The explicit expressions for the boost and rotation generators are
\begin{equation}
\mathcal{K}_{x}=\frac{1}{2}
\left(\begin{matrix}
0 & 1 \\
-1& 0
\end{matrix}\right),\hspace{0.5cm}
\mathcal{K}_{y}=\frac{1}{2}
\left(\begin{matrix}
0 & -i \\
i & 0
\end{matrix}\right),\hspace{0.5cm}
\mathcal{J}_{z}=\frac{1}{2}\left(\begin{matrix}
1 & 0 \\
0 &-1 
\end{matrix}\right).
\end{equation}

For $\psi^{(\sigma)}(x)$, where $\sigma=\pm\frac{1}{2}$ are the eigenvalues of $\mathcal{J}_{z}$, the expansion coefficients at rest, with the appropriate normalisation, are taken to be
\begin{subequations}
\begin{equation}
u(\0,{\textstyle{\frac{1}{2}}})=\sqrt{m}\left(
\begin{matrix}
1 \\
0
\end{matrix}\right),\hspace{0.5cm}
u(\0,-{\textstyle{\frac{1}{2}}})=\sqrt{m}\left(
\begin{matrix}
0 \\
1
\end{matrix}\right),
\end{equation}
\begin{equation}
v(\0,{\textstyle{\frac{1}{2}}})=\sqrt{m}\left(\begin{matrix}
0 \\
1
\end{matrix}\right),\hspace{0.5cm}
v(\0,-{\textstyle{\frac{1}{2}}})=\sqrt{m}\left(\begin{matrix}
1 \\
0
\end{matrix}\right).
\end{equation}
\end{subequations}

Having obtained the solutions of the expansion coefficients, it is straightforward to derive the field equation. Towards this end, we start with the spin-sums
\begin{equation}
\sum_{\sigma}u(\p,\sigma)\bar{u}(\p,\sigma)=
\sum_{\sigma}v(\p,\sigma)\bar{v}(\p,\sigma)=\alpha^{\mu}p_{\mu} \label{eq:spin_sums}
\end{equation}
where
\begin{equation}
\bar{u}(\p,\sigma)=u^{\dag}(\p,\sigma)\Gamma,\hspace{0.5cm}
\bar{v}(\p,\sigma)=v^{\dag}(\p,\sigma)\Gamma
\end{equation}
are the dual coefficients with $\Gamma=\alpha^{0}$ being the metric for the spin-half representation. They are defined such that the inner-products of the expansion coefficients are Lorentz-invariant. Explicit computation yields
\begin{subequations}
\begin{equation}
\bar{u}(\p,\sigma)u(\p,\sigma')=-(-1)^{1/2+\sigma}m\delta_{\sigma\sigma'},
\end{equation}
\begin{equation}
\bar{v}(\p,\sigma)v(\p,\sigma')=-(-1)^{1/2-\sigma}m\delta_{\sigma\sigma'}.\label{eq:inner_product}
\end{equation}
\end{subequations}
Multiply the spin-sums by $u(\p,\sigma)$ and $v(\p,\sigma)$ and using the orthonormal relations, we obtain
\begin{subequations}
\begin{equation}
\left[\alpha^{\mu}p_{\mu}+(-1)^{1/2+\sigma}mI\right]u(\p,\sigma)=0,
\end{equation}
\begin{equation}
\left[\alpha^{\mu}p_{\mu}+(-1)^{1/2-\sigma}mI\right]v(\p,\sigma)=0.
\end{equation}
\end{subequations}
Therefore, the field equation for $\psi^{(\sigma)}(x)$ is~\cite{Jackiw:1990ka,Cortes:1992ic,Grignani:1995vn,Horvathy:2010vm}
\begin{equation}
\left[i\alpha^{\mu}\partial_{\mu}-(-1)^{1/2-\sigma}mI\right]\psi^{(\sigma)}(x)=0. \label{eq:two_component_eq}
\end{equation}

Here it is instructive to note that since the representations are all one-dimensional, we can also derive the field equations by simply using the tensor product between the spinors and their duals within the same representation labelled by $\sigma$ from the following identities
\begin{subequations}
\begin{equation}
u(\p,\sigma)\bar{u}(\p,\sigma)=\frac{1}{2}[\alpha^{\mu}p_{\mu}-(-1)^{1/2+\sigma}mI],
\end{equation}
\begin{equation}
v(\p,\sigma)\bar{v}(\p,\sigma)=\frac{1}{2}[\alpha^{\mu}p_{\mu}-(-1)^{1/2-\sigma}mI]. \label{eq:tensor_prod}
\end{equation}
\end{subequations}
In fact, since the field operator $\psi^{(\sigma)}(x)$ is only associated with the particle state $|p,\sigma\rangle$, these identities are more relevant than the spin-sums given in eq.~(\ref{eq:spin_sums}). Additionally,  they also determine the propagator of $\psi^{(\sigma)}(x)$. A direct computation of the propagator yields
\begin{eqnarray}
S^{(\sigma)}(x,y)&=&\langle\,\,|T[\psi^{(\sigma)}(x)\overline{\psi}^{(\sigma)}(y)]|\,\,\rangle\nonumber\\
&=&\frac{i}{2}\int\frac{d^{3}p}{(2\pi)^{3}}e^{-ip\cdot(x-y)}
\left[\frac{\alpha^{\mu}p_{\mu}-(-1)^{1/2+\sigma}mI}{p\cdot p-m^{2}+i\epsilon}\right].\label{eq:prop}
\end{eqnarray}
 The Lagrangian density for $\psi^{(\sigma)}(x)$ is simply
\begin{equation}
\mathscr{L}^{(\sigma)}=\overline{\psi}^{(\sigma)}\left[i\alpha^{\mu}\partial_{\mu}
-(-1)^{1/2-\sigma}mI\right]
\psi^{(\sigma)}.
\end{equation}
While the spin-half tensor products of spinors take a relatively simple form, as we shall see, this is not the case for their higher-spin counterpart which is why the spin-sums still play an important role in deriving the field equations.
The theory of spin-half fermions and their gauge interactions have many fascinating properties that are absent from its counterparts in 3+1 dimensions. But as our main purpose is on higher-spin fermions, we will not discuss them here. For more details on classic topics such as chiral symmetry breaking and Chern-Simons theory please see~\cite{Redlich:1983dv,Redlich:1983kn,Niemi:1983rq,Appelquist:1986fd,Dunne:1998qy}.

\subsection{Higher-spin fermions}

In 2+1 dimensions, the higher-spin fermionic fields may be constructed by noting that there exists a set of general solutions of the Lorentz generators
\begin{equation}
\mathcal{K}_{x}=i J_{y},\hspace{0.5cm}
\mathcal{K}_{y}=-i J_{x},\hspace{0.5cm}
\mathcal{J}_{z}=J_{z} \label{eq:generators}
\end{equation}
where $J_{i}$, $i=x,y,z$ are the generators of the $\mathfrak{su}(2)$ algebra whose solutions are given by
\begin{subequations}
\begin{equation}
(J_{x}\pm iJ_{y})_{\ell\bar{\ell}}= \delta_{\ell,\bar{\ell}\pm1}\sqrt{(j\mp\bar{\ell})(j\pm\bar{\ell}+1)},
\end{equation}
\begin{equation}
(J_{z})_{\ell\bar{\ell}}=\ell\delta_{\ell\bar{\ell}}\label{eq:r3}
\end{equation}
\end{subequations}
with $\ell,\bar{\ell}=-j,\cdots, j$. Another possible solution for the Lorentz generators are $\mathcal{K}_{x}=-i J_{y}$ and $\mathcal{K}_{y}=i J_{x}$ and $\mathcal{J}_{z}=J_{z}$ but this is just boosting in the opposite direction with respect to $\boldsymbol{\mathcal{K}}$ given in eq.~(\ref{eq:generators}). In 2+1 dimensions, these two solutions are related to each other by a $\pi$ rotation about the $z$-axis and are therefore physically equivalent.

As we have shown in the previous section, Lorentz symmetry requires the expansion coefficients at rest to be eigenfunctions of $\mathcal{J}_{z}$. From eq.~(\ref{eq:r3}), the general solutions to $u(\0,\sigma)$ and $v(\0,\sigma)$, with appropriate normalisation are
\begin{equation}
u_{\ell}(\0,\sigma)=m^{j}\delta_{\ell\sigma},\hspace{0.5cm}
v_{\ell}(\0,\sigma)=m^{j}\delta_{\ell,-\sigma}.
\end{equation}
We are now ready to derive the field equation. The first task is to find the general metric $\Gamma$ that leaves the inner-product between the expansion coefficient for any spin Lorentz-invariant. The constraints on $\Gamma$ are
\begin{equation}
\{\Gamma,\boldsymbol{\mathcal{K}}\}=O,\hspace{0.5cm}
[\Gamma,\mathcal{J}_{z}]=O. \label{eq:cons}
\end{equation}
Substituting the Lorentz generators into eq.~(\ref{eq:cons}), the solution, up to a constant which we set to unity, is
\begin{equation}
\Gamma_{\ell\bar{\ell}}=-(-1)^{j+\bar{\ell}}\delta_{\ell{\bar{\ell}}}.
\end{equation}
Therefore, the norms of the expansion coefficients are
\begin{subequations}
\begin{equation}
\overline{u}(\p,\sigma)u(\p,\sigma)=-m^{2j}(-1)^{j+\sigma},
\end{equation}
\begin{equation}
\overline{v}(\p,\sigma)v(\p,\sigma)=-m^{2j}(-1)^{j-\sigma}.
\end{equation}
\end{subequations}

The spin-sums for $u(\p,\sigma)$ and $v(\p,\sigma)$ evaluate to
\begin{eqnarray}
\sum_{\sigma}u(\p,\sigma)\overline{u}(\p,\sigma)&=&\sum_{\sigma}v(\p,\sigma)\overline{v}(\p,\sigma)
\nonumber\\
&=&m^{2j}\exp(-2i\boldsymbol{\mathcal{K}}\cdot\bv)\Gamma\nonumber\\
&=&m^{2j}[\cosh(2\boldsymbol{\mathcal{J}\cdot\hat{\rho}})
+\sinh(2\boldsymbol{\mathcal{J}\cdot\hat{\rho}})]\Gamma\label{eq:spin_sum}
\end{eqnarray}
where $\boldsymbol{\hat{\rho}}=(-p_{y},p_{x})/|\p|$. It follows from a theorem proven by Weinberg that the right-hand side of eq.~(\ref{eq:spin_sum}) can be expressed as~\cite[app.~A]{Weinberg:1964cn}
\begin{equation}
\sum_{\sigma}u(\p,\sigma)\overline{u}(\p,\sigma)=\alpha^{\mu_{1}\cdots\mu_{2j}}p_{\mu_{1}}\cdots p_{\mu_{2j}}\label{eq:spin_sum2}
\end{equation}
where $\alpha^{\mu_{1}\cdots\mu_{2j}}$ is a symmetric and traceless tensor of rank-$2j$. To explicitly evaluate eq.~(\ref{eq:spin_sum}), we use the fact that for $j=\frac{1}{2},\frac{3}{2},\cdots$, the hyperbolic functions have the following expansions~\cite[app.~A]{Weinberg:1964cn}
\begin{subequations}
\begin{equation}
\cosh(\eta\varphi)=\cosh\varphi\left[I+\sum_{n=1}^{j-1/2}
\frac{(\eta^{2}-I)(\eta^{2}-3^{2}I)\cdots[\eta^{2}-(2n-1)^{2}I]}{(2n)!}\sinh^{2n}\varphi\right],
\end{equation}
\begin{equation}
\sinh(\eta\varphi)=\eta\sinh\varphi\left[I+\sum_{n=1}^{j-1/2}
\frac{(\eta^{2}-I)(\eta^{2}-3^{2}I)\cdots[\eta^{2}-(2n-1)^{2}I]}{(2n+1)!}\sinh^{2n}\varphi\right]
\end{equation}
\end{subequations}
where $\eta=2\boldsymbol{\mathcal{J}\cdot\hat{\rho}}$. Multiply eq.~(\ref{eq:spin_sum}) by the appropriate expansion coefficients, we obtain
\begin{subequations}
\begin{equation}
[\alpha^{\mu_{1}\cdots\mu_{2j}}p_{\mu_{1}}\cdots p_{\mu_{2j}}+(-1)^{j+\sigma}m^{2j}]u(\p,\sigma)=0,
\end{equation}
\begin{equation}
[\alpha^{\mu_{1}\cdots\mu_{2j}}p_{\mu_{1}}\cdots p_{\mu_{2j}}+(-1)^{j-\sigma}m^{2j}]v(\p,\sigma)=0.
\end{equation}
\end{subequations}
The field equation for massive fermionic field of any spin is therefore a straightforward generalisation to the spin-half case
\begin{equation}
[i^{2j}\alpha^{\mu_{1}\cdots\mu_{2j}}\partial_{\mu_{1}}\cdots\partial_{\mu_{2j}}
-(-1)^{j-\sigma}m^{2j}]\psi^{(\sigma)}(x)=0.\label{eq:higher_spin}
\end{equation}
Since the field equation is non-linear for $j>\frac{1}{2}$, we do not know if there exists a proper Lagrangian formulation for the higher-spin fields. Nevertheless, the propagators remain well-defined via the vacuum expectation value of the time-order product. However, this time, the situation is more complicated because the tensor product between expansion coefficients  are non-linear in momentum. Since we know the spin-sum, it is possible to expand the tensor product as
\begin{subequations}
\begin{equation}
u(\p,\sigma)\overline{u}(\p,\sigma)=\frac{1}{2j+1}\alpha^{\mu_{1}\cdots\mu_{2j}}
p_{\mu_{1}}\cdots p_{\mu_{2j}}+N(\p,\sigma;j),\label{eq:tensor}
\end{equation}
\begin{equation}
v(\p,\sigma)\overline{v}(\p,\sigma)=\frac{1}{2j+1}\alpha^{\mu_{1}\cdots\mu_{2j}}
p_{\mu_{1}}\cdots p_{\mu_{2j}}+N(\p,-\sigma;j)\label{eq:tensor_product}
\end{equation}
\end{subequations}
where $N(\p,\sigma;j)$ satisfies the condition
\begin{equation}
\sum_{\sigma}N(\p,\sigma;j)=O.
\end{equation}
But it is important to note that this expansion is a choice of convenience and is not unique. Generally, there is a freedom to rescale the expansion coefficients $u(\p,\sigma)$ and $v(\p,\sigma)$ since it does not change the field equation. Of course, after rescaling, the spin-sum may no longer be satisfied but this is fine as it was only used as an intermediate step to derive the field equation and do not have much significance beyond it. We shall now consider the spin-three-half field as an explicit example.



Using eqs.~(\ref{eq:tensor}) and (\ref{eq:tensor_product}), the tensor products needed to compute the propagator and the locality anti-commutator for the spin-three-half fields $\psi^{(\sigma)}(x)$, $\sigma=-\frac{3}{2},\cdots,\frac{3}{2}$ are
\begin{subequations}
\begin{equation}
u(\p,\sigma)\overline{u}(\p,\sigma)=\frac{1}{4}\alpha^{\mu\nu\rho}p_{\mu}p_{\nu}p_{\rho}+N(\p,\sigma;\textstyle{\frac{3}{2}}),
\end{equation}
\begin{equation}
v(\p,\sigma)\overline{v}(\p,\sigma)=\frac{1}{4}\alpha^{\mu\nu\rho}p_{\mu}p_{\nu}p_{\rho}+N(\p,-\sigma;\textstyle{\frac{3}{2}}).
\end{equation}
\end{subequations}
We find that although the matrices $N(\p,\sigma;\pm\textstyle{\frac{3}{2}})$ are more complicated then the spin-half case, they do in fact have an underlying structure. These matrices can be written compactly as
\begin{subequations}
\begin{equation}
N(\p,{\textstyle{\frac{3}{2}}};{\textstyle{\frac{3}{2}}})=\frac{1}{8}
[\beta_{\mu\nu\rho}p^{\mu}p^{\nu}p^{\rho}+m\beta_{\mu\nu}p^{\mu}p^{\nu}+m^{2}\beta_{\mu}p^{\mu}
+m^{3}\beta^{(3/2)}],\label{eq:1}
\end{equation}
\begin{equation}
N(\p,{\textstyle{\frac{1}{2}}};{\textstyle{\frac{3}{2}}})=\frac{1}{8}
[-\beta_{\mu\nu\rho}p^{\mu}p^{\nu}p^{\rho}+m\beta_{\mu\nu}p^{\mu}p^{\nu}-m^{2}\beta_{\mu}p^{\mu}
+m^{3}\beta^{(1/2)}],
\end{equation}
\begin{equation}
N(\p,{-\textstyle{\frac{1}{2}}};{\textstyle{\frac{3}{2}}})=\frac{1}{8}
[-\beta_{\mu\nu\rho}p^{\mu}p^{\nu}p^{\rho}-m\beta_{\mu\nu}p^{\mu}p^{\nu}-m^{2}\beta_{\mu}p^{\mu}
-m^{3}\beta^{(1/2)}],
\end{equation}
\begin{equation}
N(\p,{-\textstyle{\frac{3}{2}}};{\textstyle{\frac{3}{2}}})=\frac{1}{8}
[\beta_{\mu\nu\rho}p^{\mu}p^{\nu}p^{\rho}-m\beta_{\mu\nu}p^{\mu}p^{\nu}+m^{2}\beta_{\mu}p^{\mu}
-m^{3}\beta^{(3/2)}].\label{eq:4}
\end{equation}
\end{subequations}
The expressions for $\beta^{\mu}$ and $\beta^{(1/2,3/2)}$ are given in app.~\ref{B} but 
we will not provide expressions for the remaining the $\beta$ matrices as they are not important for our discussion.  What is important is that these matrices have precisely the structure required that make the fields local, namely
\begin{equation}
\{\psi^{(\sigma)}(x),\overline{\psi}^{(\sigma')}(y)\}=O,\hspace{0.5cm}(x-y)^{2}<0
\end{equation}
for all $\sigma=-\frac{3}{2},\cdots\frac{3}{2}$.

The notations for $N(\p,\sigma;\textstyle{\frac{3}{2}})$ can be misleading since it suggests that it is Lorentz-covariant but this is not the case. Explicit calculation shows that neither $\beta^{(1/2,3/2)}$ or $\beta^{\mu}$ transform as a scalar or a four-vector so $N(\p,\sigma;\textstyle{\frac{3}{2}})$ cannot be covariant. This means that although we can always express 
$\mathcal{D}(\Lambda)N(\p,\sigma;\textstyle{\frac{3}{2}})\mathcal{D}^{-1}(\Lambda)$ in terms of a another set of $\tilde{\beta}$ matrices, they are not related to the $\beta$ matrices by a Lorentz transformation.

Since $N(\p,\sigma;\textstyle{\frac{3}{2}})$ is not Lorentz-covariant, it cannot be a part of the propagator which must be Lorentz-covariant. Fortunately, the non-covariant terms can always be cancelled by adding the appropriate contact interacting Hamiltonians thus leaving us with a covariant propagator~\cite{Weinberg:1964cn}. However, as we have noted earlier, generally we only know that the tensor products are sums of covariant and non-covariant terms and that these terms are not unique. Therefore, additional condition must be imposed if we are to determine the propagator. Here we demand that the propagator should be a generalisation of eq.~(\ref{eq:prop}). In other words, we are expanding the tensor product as
\begin{subequations}
\begin{equation}
u(\p,\sigma)\overline{u}(\p,\sigma)=\frac{1}{4}[\alpha^{\mu\nu\rho}p_{\mu}p_{\nu}p_{\rho}-(-1)^{3/2+\sigma}m^{3}I]
+\tilde{N}(\p,\sigma;\textstyle{\frac{3}{2}}), \label{eq:e1}
\end{equation}
\begin{equation}
v(\p,\sigma)\overline{v}(\p,\sigma)=
\frac{1}{4}[\alpha^{\mu\nu\rho}p_{\mu}p_{\nu}p_{\rho}-(-1)^{3/2-\sigma}m^{3}I]
+\tilde{N}(\p,-\sigma;\textstyle{\frac{3}{2}}) \label{eq:e2}
\end{equation}
\end{subequations}
where $\tilde{N}(\p,\sigma;\pm\textstyle{\frac{3}{2}})$ are non-covariant matrices to be cancelled by the contact Hamiltonian. Therefore, the resulting covariant spin-three-half propagators take the form
\begin{equation}
S^{(\sigma)}(x,y)
=\frac{i}{4}\int\frac{d^{3}p}{(2\pi)^{3}}e^{-ip\cdot(x-y)}\left[\frac{\alpha^{\mu\nu\rho}p_{\mu}p_{\nu}p_{\rho}-(-1)^{3/2+\sigma}m^{3}I}{p\cdot p-m^{2}+i\epsilon}\right].
\end{equation}
Things would have been easier had we started with eqs.~(\ref{eq:e1}) and (\ref{eq:e2}) to derive the field equation, locality structure and propagator. The reason we chose a longer route is to emphasise the fact that the expansion of the tensor product is not unique.

We see that the higher-spin fermionic fields are more complicated due to their non-linear nature. Nevertheless, from the spin-three-half construction, it suggests that these higher-spin fermionic fields still seem to have an underlying structure. We conjecture that such structure exists for fermionic fields of any spin. That is, the matrix $N(\p,\sigma;j)$ is a generalisation of eqs.~(\ref{eq:1}-\ref{eq:4}). For example, this means that $N(\p,\pm j;j)$ takes the form
\begin{subequations}
\begin{equation}
N(\p,j;j)=\frac{1}{2(2j+1)}[\beta_{\mu_{1}\cdots\mu_{2j}}p^{\mu_{1}}\cdots p^{\mu_{2j}}
+m\beta_{\mu_{1}\cdots\mu_{2j-1}}p^{\mu_{1}}\cdots p^{\mu_{2j-1}}+\cdots
+m^{2j}\beta^{(j)}],
\end{equation}
\begin{equation}
N(\p,-j;j)=\frac{1}{2(2j+1)}[\beta_{\mu_{1}\cdots\mu_{2j}}p^{\mu_{1}}\cdots p^{\mu_{2j}}
-m\beta_{\mu_{1}\cdots\mu_{2j-1}}p^{\mu_{1}}\cdots p^{\mu_{2j-1}}+\cdots
-m^{2j}\beta^{(j)}]
\end{equation}
\end{subequations}
The form of the expansions guarantee that the locality is preserved for all fields
\begin{equation}
\{\psi^{(\sigma)}(t,\x),\overline{\psi}^{(\sigma')}(t,\mathbf{y})\}=O,\hspace{0.5cm}(x-y)^{2}<0.
\end{equation}
Since we have demanded the propagator to be a generalisation of eq.~(\ref{eq:prop}), we get
\begin{equation}
S^{(\sigma)}(x,y)
=\frac{i}{2j+1}\int\frac{d^{3}p}{(2\pi)^{3}}e^{-ip\cdot(x-y)}\left[\frac{\alpha^{\mu_{1}\cdots\mu_{2j}}p_{\mu_{1}}\cdots p_{\mu_{2j}}-(-1)^{j+\sigma}m^{2j}I}{p\cdot p-m^{2}+i\epsilon}\right].
\end{equation}

In summary, we have systematically constructed massive fermionic fields of any spin in 2+1 dimensions. The higher-spin field equations are generalisations of the spin-half Dirac equation. But due to their non-linear nature, we do not know if there exists a Lagrangian formulation or if there exists conjugate momenta that satisfy the canonical anti-commutation relations. From the propagator, we see that a spin-$j$ fermionic field has mass dimension $j+\frac{1}{2}$. This means that for $j>\frac{1}{2}$, by the power-counting argument, there are no renormalisable interactions. Nevertheless, it would be interesting to see if it is indeed possible to find Lagrangians for higher-spin fermionic fields that give us eq.~(\ref{eq:higher_spin}) and a positive-definite free Hamiltonian. This would at least allow us to study their effective interactions to see whether they inherit some of the fascinating properties of the spin-half field theory.

\section*{Acknowledgement}
This research is supported by CNPq grant 313285/2013-6. 

\appendix

\section{The $\alpha_{\mu\nu\rho}$ matrix}\label{A}
The matrices $\alpha_{\mu\nu\rho}$ are symmetric traceless tensors of rank 3 and are given by
\begin{subequations}
\begin{equation}
\alpha_{000}=\left(\begin{matrix}
1 & 0 & 0 & 0 \\
0 &-1 & 0 & 0 \\
0 & 0 & 1 & 0 \\
0 & 0 & 0 & -1
\end{matrix}\right),\hspace{0.5cm}
\alpha_{001}=\frac{1}{\sqrt{3}}\left(\begin{matrix}
0 & i & 0 & 0 \\
i & 0 & -2i/\sqrt{3} & 0  \\
0 & -2i/\sqrt{3} & 0 & i \\
0 & 0 & i & 0
\end{matrix}\right),
\end{equation}
\begin{equation}
\alpha_{002}=\frac{1}{\sqrt{3}}
\left(\begin{matrix}
0 & 1 & 0 & 0\\
-1& 0 & -2/\sqrt{3} & 0 \\
0 & 2/\sqrt{3} & 0 & 1\\
0 & 0 & -1 & 0
\end{matrix}\right),\hspace{0.5cm}
\alpha_{011}=\frac{1}{\sqrt{3}}
\left(\begin{matrix}
0 & 0 & -1 & 0\\
0 & -2/\sqrt{3} & 0 & 1 \\
-1 & 0 & 2/\sqrt{3} & 0 \\
0 & 1 & 0 & 0
\end{matrix}\right),
\end{equation}
\begin{equation}
\alpha_{012}=\frac{1}{\sqrt{3}}
\left(\begin{matrix}
0 & 0 & i & 0 \\
0 & 0 & 0 &-i \\
-i& 0 & 0 & 0 \\
0 & i & 0 & 0
\end{matrix}\right),\hspace{0.5cm}
\alpha_{022}=\frac{1}{\sqrt{3}}
\left(\begin{matrix}
0 & 0 & 1 & 0\\
0 & -2/\sqrt{3} & 0 & -1 \\
1 & 0 & 2/\sqrt{3} & 0 \\
0 & -1 & 0 & 0
\end{matrix}\right),
\end{equation}
\begin{equation}
\alpha_{111}=\left(\begin{matrix}
0 & 0 & 0 &-i \\
0 & 0 &-i & 0 \\
0 &-i & 0 & 0 \\
-i& 0 & 0 & 0
\end{matrix}\right),\hspace{0.5cm}
\alpha_{112}=\left(\begin{matrix}
0 & 0 & 0 &-1 \\
0 & 0 & -1/3 & 0 \\
0 & 1/3 & 0 & 0 \\
1 & 0 & 0 & 0
\end{matrix}\right),
\end{equation}
\begin{equation}
\alpha_{122}=\left(\begin{matrix}
0 & 0 & 0 & i \\
0 & 0 & -i/3 & 0 \\
0 & -i/3 & 0 & 0 \\
i & 0 & 0 & 0
\end{matrix}\right),\hspace{0.5cm}
\alpha_{222}=\left(\begin{matrix}
0 & 0 & 0 & 1 \\
0 & 0 &-1 & 0 \\
0 & 1 & 0 & 0 \\
-1& 0 & 0 & 0
\end{matrix}\right).
\end{equation}
\end{subequations}

\section{The $\beta^{\mu}$ and $\beta^{(1/2,3/2)}$ matrices} \label{B}
The $\beta^{\mu}$ and $\beta^{(1/2,3/2)}$ matrices given in eqs.~(\ref{eq:1}-\ref{eq:4}) are
\begin{subequations}
\begin{equation}
\beta^{0}=\left(\begin{matrix}
3 & 0 & 0 & 0 \\
0 & -1 & 0 & 0 \\
0 & 0 & 1 & 0 \\
0 & 0 & 0 & -3
\end{matrix}\right),\hspace{0.5cm}
\beta^{1}=\left(\begin{matrix}
0 & -\sqrt{3}i & 0 & 0 \\
-\sqrt{3}i & 0 & -i & 0 \\
0 & -i & 0 & -i\sqrt{3} \\
0 & 0 & -\sqrt{3}i & 0
\end{matrix}\right)
\end{equation}
\begin{equation}
\beta^{2}=\left(\begin{matrix}
0 & -\sqrt{3} & 0 & 0 \\
\sqrt{3} & 0 &-1  &0 \\
0 & 1 & 0 & -\sqrt{3} \\
0 & 0 & \sqrt{3} & 0
\end{matrix}\right),\hspace{0.5cm}
\beta^{(3/2)}=\left(\begin{matrix}
1 & 0 & 0 & 0\\
0 & 3 & 0 & 0 \\
0 & 0 & 3 & 0\\
0 & 0 & 0 & 1
\end{matrix}\right)
\end{equation}
\begin{equation}
\beta^{(1/2)}=\left(\begin{matrix}
-3 & 0 & 0 & 0\\
0 &-1 & 0 & 0 \\
0 & 0 & -1 & 0\\
0 & 0 & 0 &-3
\end{matrix}\right)
\end{equation}
\end{subequations}

\label{Bibliography}
\bibliographystyle{JHEP}  
\bibliography{references}  
\end{document}